\documentclass[aps,prd,groupedaddress,showpacs,amsmath,amssymb]{revtex4}

\usepackage{bm}
\usepackage{latexsym}
\usepackage{dcolumn}
\usepackage{amsfonts,amssymb}
\usepackage{graphicx,epsfig}
\usepackage{psfrag} 
\usepackage{amsmath} 

\def\beq{\begin{equation}}
\def\eeq{\end{equation}}
\def\br{\begin{eqnarray}}
\def\er{\end{eqnarray}}
\def\benu{\begin{enumerate}}
\def\eenu{\end{enumerate}}
\def\nn{\nonumber} 

\def\pa{{\partial}}
\def\l{\left}
\def\r{\right}

\reversemarginpar

\begin{document}
\preprint{Preprint number}

\title{Entropy of BTZ black strings in the brick wall approach} 
\author{H.~K.~Jassal}\email[]{E-mail: hkj@mri.ernet.in}
\author{L.~Sriramkumar}\email[]{E-mail: sriram@mri.ernet.in}
\affiliation{Harish-Chandra Research Institute, Chhatnag Road,
Jhunsi, Allahabad 211 019, India}
\date{\today}

%%%%%%%%%%%%%%%%%%%%%%%%%%%ABSTRACT%%%%%%%%%%%%%%%%%%%%%%%%%%%%%%%%%%%%%%%%%%

\begin{abstract}
During the last few years, exact solutions that describe black holes 
that are bound to a two-brane in a four dimensional anti-de Sitter 
bulk have been constructed.
In situations wherein there is a negative cosmological constant on
the brane, for large masses, these solutions are exactly the 
rotating BTZ black holes on the brane and, in fact, describe rotating 
BTZ black strings in the bulk. 
We evaluate the canonical entropy of a free and massless scalar field 
(at the Hawking temperature) around the rotating BTZ black string using 
the brick wall model.
We explicitly show that the Bekenstein-Hawking `area' law is satisfied 
{\it both on the brane and in the bulk}.\/
\end{abstract}
\maketitle 

%%%%%%%%%%%%%%%%%%%%%%%%%%%%%%%%%%%%%%%%%%%%%%%%%%%%%%%%%%%%%%%%%%%%%%%%%%%%%

\section{The brick wall approach to black hole entropy}\label{sec:bwbhe}

It has now been three decades since it was originally suggested 
by Bekenstein that black holes carry an entropy ${\cal S}$ that 
is proportional to the surface `area' ${\cal A}$ of their event 
horizon~\cite{bekensteinbhe}.
Soon after Bekenstein's suggestion, Hawking discovered that when 
one considers the evolution of quantum fields around black holes, 
they indeed radiate thermally~\cite{hawkingbhe}.  
Moreover, the temperature of the thermal radiation, referred to as 
the Hawking temperature, determines the exact relation between the 
`area' of the black holes and their entropy, viz. that~\cite{hawkingbhe}
\beq
{\cal S}=\l(\frac{\cal A}{4}\r)\label{eq:bhalaw}
\eeq
with the area to be `measured' in units of square of the Planck length.
This Bekenstein-Hawking `area' law is expected to apply to {\it all}\/ 
black hole solutions of the Einstein's equations.

Ever since Bekenstein's suggestion and Hawking's discovery, a variety
of approaches have been proposed to understand the microscopic origin 
of black hole entropy (for a discussion on the different approaches, 
see, for example, Refs.~\cite{ebheda} and references therein).
One of these approaches has been the semi-classical approach originally
due to 't~Hooft~\cite{thooft85}, often referred to as the brick wall model.
In this approach, the black hole geometry is assumed to be a fixed
classical background in which quantum fields propagate and the entropy 
of the black hole supposedly arises due to the statistical entropy of 
thermal fields outside the black hole horizon, evaluated in the WKB 
approximation.
But, due to the infinite blue shifting of the modes in the vicinity 
of the black hole horizon, the density of states of the matter fields 
diverge and, hence, for this model to be viable, it is necessary to 
introduce a cut-off (called the brick wall and, obviously, the reason 
behind the name of the approach) above the horizon by hand.   
This approach has been very popular in the literature, and its popularity 
can be gauged by the fact that the Bekenstein-Hawking `area'
law~(\ref{eq:bhalaw}) has been recovered for a variety of black hole
horizons (and also for accelerated as well as cosmological horizons) 
in different dimensions (for an incomplete list, see 
Refs.~\cite{bwm,kkps97,hk98}).
Though popular, the brick wall approach has its drawbacks.
Foremost amongst them being the fact that the cut-off has to be chosen 
by hand to be of the order of the Planck length in order to recover the
`area' law~(\ref{eq:bhalaw}).
However, it should be pointed out here that there have been proposals 
in the literature wherein the infinities that arise can be absorbed in 
the renormalization of the Newton's constant (in this context, see, for 
instance, Refs.~\cite{rnG}). 
Another notable limitation arises due to the fact that the entropy 
obtained depends on the number of matter fields that one chooses to 
consider.

During the last few years, scenarios have been proposed wherein the 
observable universe is considered to be a three-brane embedded in a 
higher-dimensional bulk. 
Of particular interest have been scenarios wherein the three-brane is 
embedded in a five-dimensional anti-de Sitter bulk~\cite{rs99}.
These proposals have generated considerable interest regarding the 
physics of black holes in these scenarios.
Quite a few of the standard black hole solutions have been found 
to be consistent with these backgrounds and, often, it turns out that 
these solutions actually describe black strings that extend into the 
extra dimension~(see Refs.~\cite{bwbh}; in this context, also see
Refs.~\cite{bwbhs}).

With the aim of testing the brane-world scenarios in lower dimensions, 
exact solutions describing black holes that are confined to a two-brane 
in a four-dimensional anti-de Sitter bulk have also been 
constructed~\cite{ehm00a,ehm00b}.
For cases wherein there is a negative cosmological constant on the 
brane, these solutions (for large masses of the black hole) prove to 
be precisely the rotating Banados-Teitleboim-Zanelli (BTZ, hereafter) 
black holes on the brane~\cite{btzbh} and rotating BTZ black strings 
in the bulk~\cite{ehm00b}. 

The entropy of a Schwarzschild black hole that is bound to a three-brane 
in a five-dimensional anti-de Sitter bulk has recently been evaluated 
using the brick wall approach (see Ref.~\cite{kop01}; also, see 
Ref.~\cite{medved02}; for another approach analysing the same situation,
see, for example, Ref.~\cite{das02}).
The rotating BTZ black hole exhibits the essential features of rotating 
black holes in higher dimensions in the sense that it is characterized 
by mass as well as angular momentum.
Therefore, apart from providing a technically simpler framework to work 
with, studying the properties of the rotating BTZ black string will help 
us gain an insight into the properties of rotating black strings in higher 
dimensions.
Moreover, in the earlier calculation for the Schwarzschild black hole 
embedded in a five dimensionsal anti-de Sitter bulk, the brick wall 
entropy has been shown to be proportional to the area of the event horizon 
{\it only}\/ on the brane~\cite{medved02}.
However,  one needs to establish the Bekenstein-Hawking `area' law not only 
on the brane but in the bulk as well.
With these motivations, in this paper, we evaluate the canonical entropy of 
a massless scalar field (at the Hawking temperature) around the rotating BTZ 
black string using the brick wall model.
As we shall show, when the contributions due to the bulk modes and the metric 
in the bulk are properly taken into account, we are able to recover the `area' 
law both on the brane and in the bulk.

This paper is organized as follows.
In the next section, we shall briefly review the features of rotating BTZ 
black strings that are essential for our discussion that follows. 
In Section~\ref{sec:ebtzbs}, we shall evaluate the entropy of a massless 
scalar field around the rotating BTZ black string using the brick wall 
model.
In Subsection~\ref{subsec:ebtzbs}, we shall first consider the case of a 
non-rotating BTZ black string and, then, in Subsection~\ref{subsec:erbtzbs}, 
we shall extend our analysis to the rotating case. 
Finally, we shall close with Section~\ref{sec:dscssn} with a brief summary 
of our results.

Note that we shall work in units such that $G=\hbar=c=1$.
Also, the metric signature we shall adapt will be~$(-,+,+,+)$.
  
\section{The BTZ black string---Essentials} \label{sec:btzbs}

In this section (which closely follows Ref.~\cite{ehm00b}), we shall 
summarize the construction of a BTZ black string that is bound to a 
two-brane in a four dimensional anti-de Sitter bulk (adS${}_{4}$, 
hereafter) and point out its essential features. 

A solution to the Einstein's equation with a negative cosmological constant
that describes an accelerating black hole in adS${}_{4}$ is given by the
line-element~\cite{pd76} 
\beq
ds^2=\l(\frac{1}{A^2(x-y)^2}\r)\l[-H(y)\, dt^2
+ \l(\frac{dx^2}{G(x)}\r)+\l(\frac{dy^2}{H(y)}\r) 
+ G(x)\, d\phi^2\r],\label{eq:adsc}
\eeq
where the functions $G(x)$ and $H(y)$ are given by
\beq
G(x) = \l(1 + \kappa x^2 -2mA x^3\r),\quad
H(y) = \l(\lambda - \kappa y^2 +2 m A y^3\r),
\eeq
with $\kappa=0,\pm 1$, and $m$ and $A$ are parameters that are related 
to the mass and the acceleration of the black hole.
For $\lambda>-1$, the line-element~(\ref{eq:adsc}) is called the adS 
C-metric and it satisfies the relation
\beq
R_{AB}=-\l[(3/\ell_{4}^2)\, g_{AB}\r],\quad{\rm where}\quad
\ell_{4} =\l(A\sqrt{\lambda +1}\r)^{-1},\label{eq:eeell4}
\eeq
and the bulk cosmological constant is given by $\Lambda_{4}=
-(3/\ell_{4}^2)$.
When $m=0$, the line-element~(\ref{eq:adsc}) can be written in 
terms of the variables
\beq
r= \l(\frac{\sqrt{y^2+\lambda\, x^2}}{A(x-y)}\r)\quad{\rm and}\quad
\rho = \l(\frac{1+\kappa\, x^2}{y^2+\lambda\, x^2}\r)^{1/2}\label{eq:rrho}
\eeq
as follows~\cite{ejm99}:
\beq
ds^2= r^2\l[-\l(\lambda \rho^2 -\kappa\r)\, dt^2
+ \l(\frac{d\rho^2}{\lambda\rho^2-\kappa}\r)
+\rho^2\, d\phi^2\r]
+\l(\frac{dr^2}{(r/\ell_{4})^{2}-\lambda}\r).\label{eq:BTZbs}
\eeq
In order to introduce a brane into the adS${}_{4}$ bulk, we need to 
identify a surface whose extrinsic curvature is proportional to the 
intrinsic metric.
We can then cut the spacetime off at the surface and glue two copies 
of one of the sides to form the brane.
As surfaces of constant $r$ in the above line-element satisfy this 
property, any of these surfaces can be glued to a copy of itself to 
construct the two-brane.
Also, since the constant $r$ surface has a constant curvature, there 
will be a cosmological constant induced on the brane that is given by 
$\Lambda_{3}=-(\lambda/r^2)$.
Therefore, for $\lambda >0$, the geometry of the brane will be that of 
three-dimensional anti-de Sitter spacetime (i.e. adS${}_{3}$). 
If we now assume that $\kappa=1$, then the line-element~({\ref{eq:BTZbs}) 
describes a massive BTZ black hole on the brane, and a BTZ black string 
in the bulk, {\it provided}\/ we make a periodic identification of 
$\phi$ with a period, say, $(2\pi)$.

To introduce a negative cosmological constant in the bulk, 
one introduces a second brane so that the fourth dimension 
is compactified. 
A convenient choice here is to fix one brane at $x=0$ and 
the second brane at a constant $y$  position such that 
$H'(y)=0$, again a simple choice being $y=0$.
The region $x \ge 0$ and $y \ge 0$ is glued to a copy 
of itself along the boundaries $x=0$ and $y=0$. 
These boundaries correspond to $r=(1/A)$ and 
$r=(\sqrt{\lambda}/A)$ and both surfaces have positive 
intrinsic curvature and positive tensions.
The two branes are therefore chosen to be located at 
$L_1=(1/A)$ and $L_2=(\sqrt{\lambda}/A)$.

The effects of rotation can be included by starting with the following 
line-element that describes an accelerating and rotating black hole in 
adS${}_{4}$~\cite{pd76}
\br
ds^2&=&\l(\frac{1}{A^2(x-y)^2}\r)
\Biggl[-\l(\frac{H(y)}{\l(1+a^2x^2y^2\r)}\r)\, 
\l(dt+ax^2d\phi\r)^2 +\l(1+a^2x^2y^2\r)
\biggl[\l(\frac{dy^2}{H(y)}\r)
+\l(\frac{dx^2}{G(x)}\r)\biggl]\nn\\
& &\qquad\qquad\qquad\qquad\qquad\qquad\qquad\qquad
\qquad\qquad\qquad\qquad\qquad\quad
+ \l(\frac{G(x)}{(1+a^2 x^2 y^2)}\r)
\l(d\phi-a y^2 dt\r)^2\Biggl],\label{eq:adsr}
\er
where the functions $G(x)$ and $H(y)$ are now given by
\beq
G(x) = \l(1 + \kappa x^2 -2mA x^3+\lambda a^2 x^4\r),\quad
H(y) = \l(\lambda - \kappa y^2 +2 m A y^3 +a^2 y^4\r)
\eeq
and, as before, $\kappa=0,\pm 1$.
The above line-element also satisfies the relation~(\ref{eq:eeell4}) 
and the adS C-metric~(\ref{eq:adsc}) can be obtained by setting $a=0$.
When $m=0$, the line-element~(\ref{eq:adsr}) describes adS${}_{4}$ and
in terms of the variable $r$ [as defined in Eq.~(\ref{eq:rrho})] and 
the variable $\rho$ now defined as
\beq
\rho = \l(\frac{1+\kappa x^2-a^2 x^2 y^2}{y^2+\lambda x^2}\r)^{1/2},
\eeq
the line-element reduces to
\beq
ds^2 = r^2 \biggl[-\l[\lambda \rho^2 -\kappa+(a^2/\rho^2)\r]\, dt^2
+\l(\frac{d\rho^2}{\lambda\rho^2-\kappa+(a^2/\rho^2)}\r)
+\rho^2 \, \l[d\phi-(a\, dt/\rho^2)\r]^2\biggl]
+ \l(\frac{dr^2}{(r/\ell_{4})^{2}-\lambda}\r).\label{eq:rBTZbs}
\eeq
For $\lambda>0$, $\kappa=1$, and a brane located at a constant $r$, 
the line-element above describes the geometry of a rotating BTZ black 
hole on the brane and, the entire line-element, therefore, describes 
the geometry of a rotating BTZ black string.

It is useful to note here that the issue of stability of higher 
dimensional black strings can be addressed in this scenario as 
analytical solutions to both the bulk and the brane are available.
In fact, it has been shown that the configuration of two branes in the 
four dimensional adS bulk is stable as long as the minimum transverse 
size of the black string remains larger than the four dimensional adS 
scale.
When the transverse size reaches this value, the black string breaks 
up and forms localized black holes (for a detailed discussion, see 
Ref.~\cite{ehm00b}).
This geometry is therefore useful to study effects of extra dimensions
on black hole entropy as, it closely follows, in spirit, the two brane 
Randall-Sundrum scenario.
The main difference in this model from the Randall-Sundrum scenario is 
that both the branes here have a positive tension.

\section{Entropy of the BTZ black string}\label{sec:ebtzbs}

In this section, we shall evaluate the canonical entropy of a free and
massless scalar field (at the Hawking temperature) around the BTZ black
string using the brick wall approach.
In Subsection~\ref{subsec:ebtzbs}, we shall consider the case of the 
non-rotating  BTZ black string and, in Subsection~\ref{subsec:erbtzbs}, 
we shall extend our analysis to the rotating case. 

\subsection{The non-rotating case}\label{subsec:ebtzbs}

As we had mentioned in the introductory section, in the brick wall 
approach, the entropy of black holes is attributed to the statistical 
entropy of matter fields that are in thermal equilibrium with the black 
hole.
A simple choice for the matter field would be a free and massless scalar 
field, say, $\Phi$ that satisfies the wave equation 
\beq
\Box \Phi =0.\label{eq:kg}
\eeq
In the spacetime of a BTZ black string described by the 
line-element~(\ref{eq:BTZbs}), the modes of the scalar field $\Phi$ 
can be decomposed as follows:
\beq
\Phi\l(t, \rho, \phi, r\r)=\l[e^{-iEt}\, e^{im\phi}\, R_{Em\mu}(\rho)\, 
F_{\mu}(r)\r],\label{eq:Phidcmpstn}
\eeq
where the functions $R_{Em\mu}$ and $F_{\mu}$ satisfy the differential 
equations
\beq
\l(\frac{1}{\rho\, N^2(\rho)}\r) \frac{d}{d\rho}\l[\rho\, N^2(\rho) 
\l(\frac{dR_{Em\mu}}{d\rho}\r) \r] + k_\rho^2\, 
R_{Em\mu} =0\label{eq:deREmmu}
\eeq
and
\beq
\l(\frac{1}{r^3 P(r)}\r) \frac{d}{dr}\l[r^3\, P(r)
\l(\frac{dF_{\mu}}{dr}\r)\r] + k_r^2\, F_{\mu} =0,\label{eq:deFmu}
\eeq
respectively.
The quantities $N$, $P$, $k_{\rho}$ and $k_{r}$ in the above two 
differential equations are given by the expressions
\br
N^2(\rho)=\l(\lambda\, \rho^{2}-\kappa\r),& &\quad
P^2(r)=\l[(r/\ell_{4})^2-\lambda\r],\\
k_\rho^2(E,m,\mu) = \l(\frac{1}{N^4(\rho)}\r) 
\l(E^2 - \l(\frac{m N(\rho)}{\rho^2}\r)^2-\l[\mu\, N(\rho)\r]^2\r),
\label{eq:krho}
& &\quad
k_r^2(\mu)= \l(\frac{\mu}{r P(r)}\r)^2.\label{eq:kr}
\er

The free energy ${\cal F}$ of a scalar field at the inverse 
temperature $\beta$ can be written as (see, for example, 
Ref.~\cite{thooft85})
\beq
{\cal F} = \l(\frac{1}{\beta}\r)\,
\int\, dE\, \l(\frac{dg(E)}{dE}\r)\, 
{\rm ln}\l[1 - e^{-\beta E}\r],
\eeq
where $g(E)$ denotes the number of the modes of the field upto 
the energy $E$ and the integral is to be carried out over all 
allowed values of $E$. 
In general, it turns out to be a difficult task to determine the 
number of states $g(E)$ of a quantum field exactly around a black 
hole.        
(It can be evaluated exactly only in some special cases of 
two-dimensional black holes; see, for instance, Ref.~\cite{scz97}.) 
However, under certain conditions, it can be evaluated in the WKB
approximation---a procedure that is often referred to in the 
literature as the brick-wall model~\cite{thooft85}. 

Let us now introduce an infrared and an ultraviolet cutoff so that 
the scalar field $\Phi$ vanishes at $\rho=(\rho_{H}+\epsilon)$ and 
at $\rho={\cal R}$,
where $\epsilon \ll \rho_{\rm H}$ and ${\cal R} \gg  \rho_{\rm H}$, 
$\rho_{\rm H}=\sqrt{\kappa/\lambda}$ being the horizon of the black 
string, i.e.~where $N=0$. 
(As we shall see, the entropy associated with the quantum field 
diverges as $\epsilon \to 0$ and, the ultraviolet cutoff, viz. 
$\epsilon\to 0^{+}$, is referred to as the brick-wall above the 
horizon.)
In the WKB limit, for a given $m$ and $\mu$, the number 
of radial modes upto a given energy $E$ is given by 
(cf.~Refs.~\cite{thooft85,kop01,medved02})
\beq
n_{\rho}(E,m,\mu) = \l(\frac{1}{\pi}\r) 
\int\limits_{(\rho_{H}+\epsilon)}^{\cal R}\!\!\!\! d\rho\; 
{\bar k}_{\rho}(E,m,\mu),\label{eq:scqnrho}
\eeq
where ${\bar k}_{\rho}=k_{\rho}$ when $k_{\rho}^2>0$, and 
${\bar k}_{\rho}=0$ when $k_{\rho}^2\le 0$.
Similarly, the number of modes in the direction of the bulk can 
be written as
\beq
n_{r}(\mu) = \l(\frac{1}{\pi}\r) 
\int\limits_{L_{1}}^{L_{2}}dr\, {\bar k}_r(\mu),\label{eq:scqnr}
\eeq
where, as in the radial direction, ${\bar k}_{r}=k_{r}$ when 
$k_{r}^2$ is positive, and zero otherwise.
Then, in the WKB limit, the total number of states $g(E)$ of the 
field not exceeding the energy $E$ is given by
\beq
g(E)=\int d\mu \l(\frac{dn_{r}(\mu)}{d\mu}\r)
\int dm\; n_{\rho}(E,m,\mu).\label{eq:gE}
\eeq
It should be pointed out here that the spectrum of energy $E$ 
will actually be discrete due to the boundary conditions 
imposed on the field at $\rho=(\rho_{H}+\epsilon)$ and at $\rho
={\cal R}$.
However, the gap between these energy levels will be small if
${\cal R}$ is assumed to be large.
Therefore, in what follows, we shall integrate (rather than sum)
over the allowed states.

The free energy of the quantized, free and massless scalar field 
can now be written as
\beq
{\cal F}
\approx 
\l(\frac{1}{\beta}\r)  
\int dE\, \int d\mu\, \l(\!\frac{dn_{r}}{d\mu}\!\r)\,
\int dm\, \l(\!\frac{dn_{\rho}}{dE}\!\r)\, 
{\rm ln}\l[1 -  e^{-\beta E} \r]
\eeq
which, on integrating over $E$ by parts, reduces to~\cite{medved02}
\beq
{\cal F}
=-\int dE\, \int dm\, \int d\mu\, \l(\!\frac{dn_{r}}{d\mu}\!\r)
\l(\frac{n_{\rho}}{e^{\beta E}-1}\r).\label{eq:calF}
\eeq
From the expressions~(\ref{eq:kr}) and (\ref{eq:scqnr}), we find that
\beq
\l(\!\frac{dn_{r}}{d\mu}\!\r)
=\l(\frac{1}{\pi}\r)\int\limits_{L_{1}}^{L_{2}}\,
dr\, \l(\frac{d{\bar k}_{r}}{d\mu}\r)
=\l(\frac{1}{\pi}\r) \int\limits_{L_{1}}^{L_{2}} 
\frac{dr}{r\, P}.\label{eq:bf}
\eeq 
On using this result and the definition~({\ref{eq:scqnrho}) in the 
expression~(\ref{eq:calF}) for the free energy~${\cal F}$, we 
obtain that
\beq
{\cal F}=-\l(\frac{1}{\pi}\r)\, \int\limits_{L_{1}}^{L_{2}}\! 
\frac{dr}{r\, P}
\int\limits_{(\rho_{H}+\epsilon)}^{\mathcal R}\!\!\!\! 
\frac{d\rho}{N^2}\; 
\int dE\, \int d\mu\, \int dm\, \l(\frac{1}{e^{\beta E} -1}\r)
\l[E^2 - \l(m N/\rho\r)^2-\l(\mu\, N\r)^2\r]^{1/2}.
\eeq
The limits of integration on $m$ and $\mu$ are determined by the range 
of values for which the argument of the square root remains positive.
This condition leads to the following limits for the remaining integrals:
\beq
0\le m \le \l(\rho/N\r) \l[E^2 - \mu^2 N^2\r]^{1/2},\quad
0\le \mu \le \l(E/N\r)\quad{\rm and}\quad 
0\le E\le \infty.
\eeq
On carrying out the integrals over $m$ and $\mu$, in that order, 
we find that the expression for ${\cal F}$ simplifies to a product
of two separate integrals which in turn can be easily integrated to
yield the result
\beq
{\cal F}
=-\l(\!\frac{1}{6\, \pi}\!\r)\, \int\limits_{L_{1}}^{L_{2}} 
\frac{dr}{r\, P}
\int\limits_{(\rho_{H}+\epsilon)}^{\cal R}\!\!\!\!
\frac{d\rho\; \rho}{N^4}\, 
\int\limits_0^{\infty} \frac{dE\, E^3}{\l(e^{\beta E} -1\r)}
= \l(\!\frac{\zeta(4)}{2\, \lambda^{2}\, \beta^{4}}\!\r)\,
\int\limits_{L_{1}}^{L_{2}} \frac{dr}{\l[r\, P(r)\r]}\,
\l[\l(\frac{1}{{\cal R}^2 -\rho_H^2}\r) 
- \l(\frac{1}{(\rho_{H} + \epsilon)^2 - \rho_{H}^2}\r)\r],
\eeq
where $\zeta(4)$ denotes the Riemann $\zeta$-function.
As ${\cal R} \to\infty$, this expression reduces to
\beq
{\cal F}
= -\l(\!\frac{\zeta(4)}{2\, \lambda^2\, \beta^{4}}\!\r)\,
\int\limits_{L_{1}}^{L_{2}} \frac{dr}{r\, P}\,
\l(\frac{1}{(\rho_{H} + \epsilon)^2 - \rho_{H}^2}\r)
\eeq
and the need for a brick-wall above the horizon is evident from this 
expression for the free energy ${\cal F}$---it would have diverged 
had $\epsilon$ been zero.
As we had mentioned in the introductory section, this divergence 
arises due to the infinite blue-shifting of the modes near the
horizon~\cite{thooft85}.

The entropy ${\cal S}$ corresponding to the free energy ${\cal F}$ is 
given by~\cite{thooft85,medved02}
\beq
{\cal S}=\beta^2 \l(\frac{\pa {\cal F}}{\pa \beta}\r)
= \l(\!\frac{2\,\zeta(4)}{\lambda^{2}\, \beta^{3}}\!\r)\,
\int\limits_{L_{1}}^{L_{2}} \frac{dr}{r\, P}\,
\l(\frac{1}{(\rho_{H} + \epsilon)^2 - \rho_{H}^2}\r).
\eeq
Since the scalar field is in thermal equilibrium with the black string,
the entropy ${\cal S}_{\rm BS}$ of the BTZ black string can be obtained 
by evaluating the above entropy at the inverse Hawking temperature 
$\beta_{\rm H}$ of the string.
We find that, the entropy ${\cal S}_{\rm BS}$ of the black string is 
given by 
\beq
{\cal S}_{BS}
= \l(\!\frac{2\,\zeta(4)}{\lambda^{2}\, \beta_{H}^{3}}\!\r)
\int\limits_{L_{1}}^{L_{2}} \frac{dr}{r\, P}\,
\l(\frac{1}{(\rho_{H} + \epsilon)^2 - \rho_{H}^2}\r)
\eeq
an expression which, clearly, diverges as $\epsilon$ approaches zero.
 
In order to do away with this divergence, we shall now follow the
't Hooft's procedure~\cite{thooft85} and express the entropy of the 
black string in terms of an invariant ultraviolet cutoff, say, 
${\tilde \epsilon}$  rather than the brick wall location $\epsilon$.
The invariant cutoff ${\tilde \epsilon}$ is defined as the invariant 
distance between the horizon and the brick wall, viz.
\beq
\tilde{\epsilon}=\int\limits_{\rho_{\rm H}}^{\rho_{\rm H} + \epsilon}
\!\!d\rho\, \sqrt{g_{\rho \rho}} 
= r\!\! \int\limits_{\rho_{\rm H}}^{\rho_{\rm H}+\epsilon}
\!\! \frac{d\rho}{N}=\l(r/\sqrt{\lambda}\r)\, 
{\rm cosh}^{-1}\l(\frac{\rho_{\rm H}+\epsilon}{\rho_{\rm H}}\r).
\label{eq:ico}
\eeq
On using this relation, we find that the entropy ${\cal S}_{\rm BS}$ 
of the black string can then be written in terms of ${\tilde \epsilon}$ 
as follows:
\beq
{\cal S}_{\rm BS}
= \l(\!\frac{\lambda\, \zeta(4)\, \rho_{H}}{4 \pi^3}\!\r)\;
\int\limits_{L_{1}}^{L_{2}} \frac{dr}{r\, P}\;
\mathrm{sinh}^{-2}\l(\sqrt{\lambda}\, {\tilde \epsilon}/r\r),
\eeq
where we have made use of the fact that $\beta_{\rm H}=(2\pi/\lambda\,
\rho_{\rm H})$.
If we now assume that ${\tilde \epsilon}$ is a very small quantity---i.e.
the brick wall is located very close to the horizon---then, in the leading
order, we have
\beq
{\cal S}_{\rm BS}
\simeq 
\l(\!\frac{\lambda\, \zeta(4)\, \rho_{\rm H}}{4\, \pi^3}\!\r)\;
\int\limits_{L_{1}}^{L_{2}} \frac{dr}{r\, P}\,
\l(\!\frac{r}{\sqrt{\lambda}\, {\tilde \epsilon}}\!\r)^{2}
=\l(\!\frac{\zeta(4)}{4 \pi^3\, {\tilde \epsilon}^2}\!\r)\, 
\l(\rho_{\rm H}\, {\cal L}\r),
\eeq
where 
\beq
{\cal L}=\int\limits_{L_{1}}^{L_{2}}\, \frac{dr\; r}{P}
\label{eq:calL}
\eeq
is the invariant length of the black string along the extra dimension.
We can now write the above expression for the entropy as
\beq
{\cal S}_{\rm BS}
=\l(\!\frac{\zeta(4)}{8 \pi^4\, 
{\tilde \epsilon}^2}\!\r)\, {\cal A}_{\rm BS},
\eeq
where ${\cal A}_{\rm BS}=\l[(2\pi\, \rho_{\rm H})\, {\cal L}\r]$ is 
the area of the black string horizon.
Evidently, the Bekenstein-Hawking area law~(\ref{eq:bhalaw}) will 
be satisfied if we choose ${\tilde \epsilon}$ to be of the order 
of Planck length.

\subsection{The case of the rotating black string}\label{subsec:erbtzbs}

In this subsection, we shall evaluate the entropy of a rotating black 
string following the method outlined in the last section.
However, in the background of a rotating black hole, the calculation of 
the entropy turns out to be a bit more involved due to the presence of 
the super-radiant modes~\cite{hk98}. 
The contribution of the super-radiant modes to the entropy of the rotating 
BTZ black string needs to be taken into account carefully and, when done so, 
as we shall see, one regains the result for the non-rotating black string 
in the limit of zero angular momentum.

Recall that the rotating BTZ black string is described by the 
line-element~(\ref{eq:rBTZbs}).
The modes of the free and massless scalar field $\Phi$ propagating 
in such background can be decomposed exactly as we did earlier in
Eq.~(\ref{eq:Phidcmpstn}) in the case of the non-rotating black 
string.
The decomposition~(\ref{eq:Phidcmpstn}) of the scalar field mode 
leads to the same differential equation for $R_{Em\mu}$ as
before [viz.~Eq.~(\ref{eq:deREmmu})], but with $N$ and $k_{\rho}$ 
now given by  
\br
N^2&=&\l[\lambda\, \rho^2 - \kappa + \l(a^{2}/\rho^2\r)\r],
\label{eq:Nrbs}\\
k_\rho^2(E,m,\mu)
&=&\l(\!\frac{1}{N^{4}(\rho)}\!\r)
\l(E^{2} - \l(\frac{m^2}{\rho^2}\r)\l(\lambda \rho^2 -\kappa\r) 
- \l(\frac{2a mE}{\rho^2}\r)-\l[\mu\, N(\rho)\r]^2\r).
\label{eq:krhors}
\er
As will be evident later, it proves to be convenient to write 
the above expression for $k_\rho^2(E,m,\mu)$ as follows:
\beq
k_\rho^2(E,m,\mu)
=\l(\!\frac{1}{N^{4}(\rho)}\!\r)
\l[\l(E-m\, \Omega_{-}\r)\, \l(E-m\, \Omega_{+}\r)
-\l[\mu\, N(\rho)\r]^2\r]^{1/2},
\eeq
where $\Omega_{\pm}=\l[\l(a\pm \rho\, N(\rho)\r)/\rho^2\r]$.
Moreover, as expected, the bulk mode $F_{\mu}$ satisfies the same 
differential equation as in the non-rotating case, viz. 
Eq.~(\ref{eq:deFmu}).
The inner and the outer horizons of the rotating black string, 
say,~$\rho_{+}$ and $\rho_{-}$, correspond to the points where 
$N$ vanishes.
They are given by
\beq
\rho_{\pm}^2 
= \l(\frac{1}{2 \lambda}\r)\, 
\l[\kappa \pm \sqrt{\kappa^2 - 4 \lambda a^2} \r]
\eeq
and it is useful to note that the quantities appearing in the 
metric can be written in terms of $\rho_{+}$ and $\rho_{-}$ as 
follows:
\beq
N^2= \l(\frac{\lambda}{\rho^2}\r)\, \l[\l(\rho^2 - \rho_+^2\r)\,
\l(\rho^2 - \rho_-^2\r)\r],\qquad
\kappa=\lambda\, \l(\rho_{+}^{2}+\rho_{-}^{2}\r)\quad{\rm and}
\quad a=\sqrt{\lambda}\, \l(\rho_{+}\, \rho_{-}\r).
\eeq

As we mentioned above, around a rotating black hole, there exist
super-radiant modes whose contribution needs to be taken into 
account carefully.
These are modes for which $\l(E-m\, \Omega_{\rm H}\r)<0$, where 
$\Omega_{\rm H}$ is the angular speed of observers with zero 
angular momentum (i.e. ZAMOS) at the outer horizon $\rho_{+}$. 
Since the angular speed of ZAMOS at a given radius is $\Omega
=\l(a/\rho^2\r)$, we have $\Omega_{\rm H}=\l(a/\rho_{+}^2\r)$.
Around the rotating black string, the free energy ${\cal F}$ of 
the scalar field $\Phi$ can be written as 
\beq
{\cal F}=\l({\cal F}_{\rm NS}+{\cal F}_{SR}\r),
\eeq
where the non-super-radiant and the super-radiant contributions,
viz. ${\cal F}_{\rm NS}$ and ${\cal F}_{SR}$, are given by (for 
a detailed discussion on this point, see Ref.~\cite{hk98})
\br
{\cal F}_{\rm NS} 
&=& \l(\frac{1}{\beta}\r)\,
\int\limits_{\notin\,{\rm SR}}\, dE\, \l(\frac{dg(E)}{dE}\r)\, 
{\rm ln}\l[1 - e^{-\beta \l(E-m\, \Omega_{\rm H}\r)}\r],
\label{eq:calFNS}\\
{\cal F}_{\rm SR} 
&=& \l(\frac{1}{\beta}\r)\,
\int\limits_{\in\, {\rm SR}}\, dE\, \l(\frac{dg(E)}{dE}\r)\, 
{\rm ln}\l[1 - e^{\beta \l(E-m\, \Omega_{\rm H}\r)}\r].
\label{eq:calFSR}
\er
In these expressions, $g(E)$, as earlier, is given by Eq.~(\ref{eq:gE}) 
with $n_{\rho}$ and $n_{r}$ defined in Eqs.~(\ref{eq:scqnrho}) 
and~(\ref{eq:scqnr}).
Also, as in the non-rotating case, ${\bar k}_{\rho}=k_{\rho}$ when 
$k_{\rho}^2$ is positive, and zero otherwise. 
However, note that, $k_{\rho}$ is now given by Eq.~(\ref{eq:krhors}),
and the brick-wall is located just beyond the outer horizon at 
$(\rho_{+} +\epsilon)$.
The limits for $\mu$ prove to be the same for the super-radiant as well 
as the non-super-radiant modes, and the limits are given by $0\, \le\, 
\mu\, \le\; \l[(E-m\,\Omega_{+})\,(E-m\,\Omega_{-})/N\r]$.

Let us first consider the contribution to the free energy due to 
the non-super-radiant modes.
It is given by
\beq
{\cal F}_{\rm NS}
\approx 
\l(\frac{1}{\beta}\r)  
\int dE\, \int d\mu\, \l(\!\frac{dn_{r}}{d\mu}\!\r)\,
\int dm\, \l(\!\frac{dn_{\rho}}{dE}\!\r)\, 
{\rm ln}\l[1 -  e^{-\beta \l(E-m\, \Omega_{\rm H}\r)}\r]
\eeq
which, on integrating over $E$ by parts, reduces to
\br
\!\!\!\!\!\!\!\!\!\!
{\cal F}_{\rm NS}\! 
&=&\! -\l(\frac{1}{\pi}\r)\!\! 
\int\limits_{\l(\rho_{+} +\epsilon\r)}^{\cal R} \!\!\!\!
\frac{d\rho}{N^2}\, 
\int dE \int d\mu \l(\!\frac{dn_{r}}{d\mu}\!\r)\!
\int dm\, \l(\!\frac{1}{e^{\beta(E -m\,\Omega_{\rm H})}-1}\!\r)\;
\l[\l(E-m\, \Omega_{-}\r)\, \l(E-m\, \Omega_{+}\r)
-\l[\mu\, N\r]^2\r]^{1/2}\nn\\
& &+ \l(\frac{1}{\pi\beta}\r)\!\! 
\int\limits_{\l(\rho_{+} +\epsilon\r)}^{\cal R}\!\!\!\! 
\frac{d\rho}{N^2}\, 
\int d\mu \l(\!\frac{dn_{r}}{d\mu}\!\r)\!
\int dm \l[\l(E-m\, \Omega_{-}\r)\, 
\l(E-m\, \Omega_{+}\r)-\l[\mu\, N\r]^2\r]^{1/2}\, 
{\rm ln}\l[1 - e^{-\beta(E-m\,
\Omega_{\rm H})}\r]_{E_{\rm min}}^{E_{\rm max}}.
\label{eq:calFNS1}
\er
It is useful to note that the second term in the above expression 
vanishes for the case of the non-rotating black string.) 
The above expression for ${\cal F}_{\rm NS}$ can be conveniently divided 
into a part with positive angular momentum states (i.e. with $m\ge 0$) 
and another with negative angular momentum states (i.e. with $m<0$). 
For $0\le m< \infty$, the limits of the integrals over $\rho$ and
$E$ are given by $\l(\rho_{+}+\epsilon\r) \le \rho \le {\cal R}$ and
$\l(m\, \Omega_{+}\r) \le E < \infty$.
Also, since $\Omega_{+} > \Omega_{-}$, we have $E>\l(m\, \Omega_{+}\r)$.
For the negative angular momentum states (i.e. when $\infty < m \le 0$), 
we find that, for $\l(\rho_{+}+\epsilon\r)<\rho<\rho_{\rm erg}
=\sqrt{k/\lambda}$, the limits of the integral over $E$ is given by 
$0 \le E \le \infty$, whereas, for $\l(\rho_{+}+\epsilon\r)\le\rho\le 
\rho_{\rm erg}$ the limits of the integral over $E$ are given by 
$\l(m\, \Omega_{-}\r) \le E < \infty$. 
Therefore, the contribution to the free energy due to the non-super-radiant 
modes with positive  and negative angular momentum states (which we shall
denote as ${\cal F}_{\rm NS}^{+}$ and ${\cal F}_{\rm NS}^{-}$, respectively) 
are given by 
\beq
{\cal F}_{\rm NS}^{+}  
= -\l(\frac{1}{\pi}\r)\int d\mu\, \left(\frac{dn_{r}}{d\mu}\right) 
\int\limits_{\l(\rho_{+} + \epsilon\r)}^{\cal R} 
d\rho\, \int\limits_{0}^{\infty}dm 
\int\limits_{\l(m\Omega_{+}\r)}^{\infty} dE
\l(\frac{{\bar k}_{\rho}}{e^{\beta(E - m \Omega_{\rm H})}-1}\r)
\eeq
and
\br
{\cal F}_{\rm NS}^{-} 
&=& -\l(\frac{1}{\pi}\r) \int d\mu\,
\l(\frac{dn_{r}}{d\mu}\r)
\int\limits_{\l(\rho_{+} + \epsilon\r)}^{\rho_{\rm erg}}
d\rho \int_{-\infty}^{0} dm  \int_{0}^{\infty} dE  
\l(\frac{{\bar k}_{\rho}}{e^{\beta(E - m\, \Omega_{\rm H})}-1}\r)\nn\\
& &\quad
+\l(\frac{1}{\pi}\r) \int d\mu\,\left(\frac{dn_{r}}{d\mu}\right)
\int\limits_{\rho_{\rm erg}}^{\cal R} d\rho  
\int\limits_{-\infty}^{0} dm  
\int\limits_{\l(m\,\Omega_{-}\r)}^{\infty}dE\,
\l(\frac{{\bar k}_{\rho}}{e^{\beta(E - m\, \Omega_{\rm H})}-1}\r)\nn\\
& &\quad
-\l(\frac{1}{\pi \beta}\r) \int d\mu\, \left(\frac{dn_{r}}{d\mu}\right) 
\int\limits_{\l(\rho_++\epsilon\r)}^{\rho_{erg}} d\rho
\int\limits_{-\infty}^{0} dm\; {\bar k}_{\rho}(E=0)\;
{\rm ln}\l(1-e^{m\,\beta\, \Omega_{\rm H}} \right).
\er
The contribution to the free energy due to the super-radiant modes can
be evaluated in a similar fashion.
In this case, for $0\le m < \infty$, the limits of integration for $\rho$ 
and $E$ turn out to be: $\l(\rho_{+} + \epsilon\r) \le \rho \le  
\rho_{\rm erg}$ and $0\le E \le \l(m\Omega_{-}\r)$. 
The corresponding contribution to the free energy is given by 
\br
{\cal F}_{\rm SR} 
&=& -\l(\frac{1}{\pi}\r)\int d\mu\, \l(\frac{dn_{r}}{d\mu}\r)
\int\limits_{\rho_++\epsilon}^{\rho_{erg}}d\rho 
\int\limits_{-\infty}^{0} dm  
\int\limits_{\l(m\Omega_{+}\r)}^{\infty} dE\;  
\l(\frac{{\bar k}_{\rho}}{e^{-\beta(E - m \Omega_{\rm H})}-1}\r)\nn\\
& &\qquad\qquad
+\l(\frac{1}{\pi \beta}\r)
\int d\mu\, \l(\frac{dn_{r}}{d\mu}\r) 
\int\limits_{\l(\rho_{+}+ \epsilon\r)}^{\rho_{\rm erg}} d\rho
\int\limits_{0}^{\infty}dm\, \bar{k}_{\rho}\; 
{\rm ln}\l(1-e^{\beta m\Omega_{\rm H}}\r).  
\er

For convenience, we shall club the contribution to ${\cal F}_{\rm NS}$ 
due to the modes with $E<\l(m\, \Omega_{-}\r)$ with the contribution 
due to the super-radiant modes~\cite{hk98}.
Moreover, for $0 \le E \le \l(m\, \Omega_{-}\r)$, when $m$ is set to 
$-m$ in ${\cal F}_{\rm NS}$, we obtain a term which is the same as the
contribution in the super-radiant modes, but with an opposite sign. 
Therefore, we shall drop these two terms hereafter.
Moreover, the second  expression in Eq.~(\ref{eq:calFNS1}) vanishes for 
$\rho_{\rm erg} \le \rho \le {\cal R}$. 
Due to these reasons, we have
\br
{\cal F}_{\rm NS}^{+}  
&=& -\l(\frac{1}{4}\r)\, \int\limits_{L_{1}}^{L_{2}}\,  
\frac{dr}{r\, P}\, 
\int\limits_{\l(\rho_{+} + \epsilon\r)}^{\cal R}\! 
\frac{d\rho}{N^{3}(\rho)}
\int_{0}^{\infty}dm 
\int\limits_{\l(m\Omega_{+}\r)}^{\infty} dE\;
\l[\frac{(E - m\Omega_{+})(E-m\Omega_{-})}{e^{\beta(E 
- m\Omega_{\rm H})}-1}\r],\\ 
{\cal F}_{\rm NS}^{-} 
&=& -\l(\frac{1}{4}\r)\, \int\limits_{L_{1}}^{L_{2}}\,  
\frac{dr}{r\, P}\,
\int\limits_{\l(\rho_{+} + \epsilon\r)}^{\rho_{\rm erg}}\!
\frac{d\rho}{N^{3}(\rho)}\,
\int\limits_{-\infty}^{0} dm\, \int\limits_{0}^{\infty} dE  
\l[\frac{(E - m\Omega_{+})(E-m\Omega_{-})}{e^{\beta(E 
- m \Omega_{\rm H})}-1}\r]\nn\\
& &\quad 
+\l(\frac{1}{4}\r)\, \int\limits_{L_{1}}^{L_{2}}\,  
\frac{dr}{r\, P}\, 
\int_{\rho_{\rm erg}}^{\cal R}\! \frac{d\rho}{N^{3}(\rho)} 
\int\limits_{-\infty}^{0} dm  
\int\limits_{\l(m\Omega_{-}\r)}^{\infty}dE\, 
\l[\frac{(E - \Omega_+ m)(E-\Omega_+ m)}{e^{\beta(E 
- m \Omega_{\rm H})}-1}\r]\nn\\ 
& &\quad
-\l(\frac{{1}}{4 \beta}\r)\, \int\limits_{L_{1}}^{L_{2}}\,  
\frac{dr}{r\, P}\, 
\int\limits_{\l(\rho_{+} \epsilon\r)}^{\rho_{\rm erg}}
\frac{d\rho}{N^{3}(\rho)} 
\int\limits_{-\infty}^{0}dm\, \l(m^2\, \Omega_{+} \Omega_{-}\r)\,
{\rm ln}\l(1-e^{\beta m\Omega_{\rm H}}\r),\\
{\cal F}_{\rm SR} 
&=& -\l(\frac{1}{4}\r)\, \int\limits_{L_{1}}^{L_{2}}\,  
\frac{dr}{r\, P}\,
\int\limits_{\rho_{+}}^{\rho_{\rm erg}}
\frac{d\rho}{N^3(\rho)}
\int_{-\infty}^{0} dm\,
\int\limits_{\l(m\Omega_{+}\r)}^{\infty} dE\, 
\l[\frac{(E - \Omega_+ m)(E-\Omega_+ m)}{e^{-\beta(E - m
\Omega_{\rm H})}-1}\r]\nn\\ 
& &\quad
+\l(\frac{1}{4\beta}\r)\, \int\limits_{L_{1}}^{L_{2}}\,  
\frac{dr}{r\, P}\,
\int\limits_{\l(\rho_{+}+ \epsilon\r)}^{\rho_{\rm erg}} 
\frac{d\rho}{N^{3}(\rho)} 
\int\limits_{0}^{\infty}dm \l(m^2\, \Omega_{+} \Omega_{-}\r)
\l[{\rm ln}\left(1-e^{-\beta m\Omega_{\rm H}}\right)\r],  
\er
where we have made use the expression~(\ref{eq:bf}) for 
$\l(dn_{r}/d\mu\r)$. 

Let us now first consider the contribution due to the positive angular 
momentum states of the the non-super-radiant modes.
The $E$ and $m$ integrals in the expression for ${\cal F}_{\rm NS}^{+}$
can be separated by the coordinate transformation $E= \l(m\Omega_{+} x\r)$ 
and we obtain that
\beq
{\cal F}_{\rm NS}^{+} 
=  -\l(\frac{1}{4}\r)\, \int\limits_{L_{1}}^{L_{2}}\,  
\frac{dr}{r\, P}\, 
\int\limits_{\l(\rho_{+} + \epsilon\r)}^{\cal R} 
\frac{d\rho}{N^{3}(\rho)}\, \Omega_{+}^2
\int\limits_{1}^{\infty} dx \l(x-1\r)\,\l(\Omega_+ x -\Omega_{-}\r)
\int\limits_{0}^{\infty}dm\, 
\l(\frac{m^3}{e^{\beta m\l(\Omega_+ x -\Omega_{\rm H}\r)} -1}\r).
\eeq
On carrying out the integral over $m$ first and then subsequently
integrating over $x$, we obtain that
\beq
{\cal F}_{\rm NS}^{+}
=-\l(\frac{\zeta(4) \Gamma(4)}{24 \beta^4}\r)\, 
\int\limits_{L_{1}}^{L_{2}}\,  \frac{dr}{r\, P}\,
\int\limits_{\l(\rho_{+} + \epsilon\r)}^{\cal R} 
\frac{d\rho}{N^3(\rho)} 
\l[\frac{3 \Omega_{+} - 2 \Omega_{\rm H} 
- \Omega_{-}}{\l(\Omega_{\rm H} -\Omega_{+}\r)^2}\r].
\eeq
Similarly, for ${\cal F}_{\rm NS}^{-}$, we can transform variables to
$E=\l(mx\r)$ and $E=\l(\Omega_{-} m x\r)$ for the first and second terms,
respectively.  
We find that we can write
\br
{\cal F}_{\rm NS}^{-}
&=&-\l(\frac{\zeta(4)\Gamma(4)}{24\beta^4}\r)\, 
\int\limits_{L_{1}}^{L_{2}}\, \frac{dr}{r\, P}\,
\int\limits_{\l(\rho_{+} + \epsilon\r)}^{\rho_{\rm erg}}
\frac{d\rho}{N^3(\rho)}
\l(\frac{2\Omega_{\rm H}^2 + 2 \Omega_{+}\Omega_{-} 
+ \l(\Omega_{+} + \Omega_{-}\r)\Omega_H}{\Omega_{\rm H}^3}\r)\nn\\
& &\quad
-\l(\frac{\zeta(4) \Gamma(4)}{24 \beta^4}\r)\,
\int\limits_{L_{1}}^{L_{2}}\, \frac{dr}{r\, P}\,
\int\limits_{\rho_{\rm erg}}^{\cal R} \frac{d\rho}{N^3(\rho)} 
\l(\frac{\Omega_+ + 2 \Omega_{\rm H} -
3 \Omega_{-}}{\l(\Omega_{\rm H} -\Omega_{+}\r)^2}\r)\nn\\
& &\quad
-\l(\frac{1}{4 \beta}\r)\, 
\int\limits_{L_{1}}^{L_{2}}\, \frac{dr}{r\, P}\, 
\int\limits_{\l(\rho_{+} + \epsilon\r)}^{\rho_{\rm erg}}
\frac{d\rho}{N^{3}(\rho)} 
\int\limits_{-\infty}^{0} dm \l(m^2\, \Omega_{+} \Omega_{-}\r)
\l[{\rm ln}\l(1-e^{\beta m\Omega_{\rm H}}\r)\r].
\er
For the contribution due to the super-radiant modes, we can make the
transformation is $E=\l(m\Omega_{-}x\r)$ and, after integrating over
the variables $m$ and $x$, we obtain that
\br
{\cal F}_{\rm SR} 
&=& -\l(\frac{\zeta(4) \Gamma(4)}{24 \beta^4}\r)\,
\int\limits_{L_{1}}^{L_{2}}\, \frac{dr}{r\, P}\,
\int\limits_{\l(\rho_{+} + \epsilon\r)}^{\rho_{\rm erg}} 
\frac{d\rho}{N^3(\rho)}
\l(\frac{\Omega_{-}^{2}\l(3\Omega_{+}\Omega_{\rm H} 
- \Omega_{-} \Omega_{\rm H} - 2 \Omega_{-}\Omega_{+}\r)}{\Omega_{\rm H}^3 
\l(\Omega_{\rm H} -\Omega_{+}\r)^2}\r)\nn\\ 
& &\quad
+\l(\frac{1}{4\,\beta}\r)\, \int\limits_{L_{1}}^{L_{2}}\,  
\frac{dr}{r\, P}\,
\int\limits_{\l(\rho_{+} + \epsilon\r)}^{\rho_{\rm erg}} 
\frac{d\rho}{N^{3}(\rho)} \int\limits_{0}^{\infty} dm\,
\l(m^2\, \Omega_{+}\Omega_{-}\r)
\l[{\rm ln}\left(1-e^{-\beta m\Omega_{\rm H}} \r)\r]. 
\er 
We need not explicitly carry out the integral containing the 
logarithmic term, as a similar term arise in the contribution 
due to the non-super-radiant modes with an opposite sign and, 
hence will not contribute in the final expression for the free 
energy. 
Morover, in evaluating the integrals in the above expressions, 
we drop terms which vanish in the limit of $\epsilon\to 0$, as 
we are interested only in the divergent part.
On retaining the leading order terms in $\epsilon$, we find that
the various contributions to the free energy are given by
\br
{\cal F}_{\rm NS}^{+}
&=& - \frac{1}{2}\l(\frac{\zeta(4)}{\beta^4}\r)\,
\int\limits_{L_{1}}^{L_{2}}\,  \frac{dr}{r\, P}\, 
\l(\frac{\rho_{+}}{\lambda^2(\rho_+^2 - \rho_-^2 )^3}\r)  
\l[\l(\frac{\rho_{+}^{3}\l(\rho_{+}^{2} 
- \rho_{-}^{2}\r)}{2\rho_{+}\epsilon}\r)
-\l(\rho_-^3 + 3 \rho_- \rho_{+}^{2}\r) 
+ 2 \rho_{-} \rho_{+}^{2}{\sqrt{2 \epsilon \rho_{+}}}\r],\\ 
{\cal F}_{\rm NS}^{-}
&=& - \frac{1}{2}\l(\frac{\zeta(4)}{\beta^4}\r)\,
\int\limits_{L_{1}}^{L_{2}}\, \frac{dr}{r\, P}\, 
\l(\frac{1}{\lambda^2(\rho_+^2 - \rho_-^2 )^3}\r)
\l[\frac{\rho_{-}}{\rho_{+}}(\rho_{-}^{4} + \rho_{-}^{2} 
\rho_{+}^{2} + 4 \rho_{+}^{4})(\rho_+^2 - \rho_-^2) 
+3\, \l(\rho_{+}\, \rho_{-}\r)^2\, \frac{\l(\rho_{+}^{2}
-\rho_{-}^{2}\r)^{3/2}}{\sqrt{2 \epsilon \rho_{+}}}\r],\quad\\
{\cal F}_{\rm SR}
&=& - \frac{1}{2} \l(\frac{\zeta(4)}{\beta^4}\r)\,
\int\limits_{L_{1}}^{L_{2}}\, \frac{dr}{r\, P}\,
\l(\frac{1}{\lambda^2(\rho_+^2 - \rho_-^2 )^3}\r)
\left[\rho_{+}^{4} + \frac{3 \rho_+^6}{\rho_-^2} 
+\frac{\rho_+^4(\rho_+^2-\rho_-^2)}{2 \rho_+ \epsilon}
-\frac{3\rho_{+}^{5}}{\rho_{-}} 
\frac{\sqrt{\rho_{+}^{2} -\rho_{-}^{2}}}
{\sqrt{2 \epsilon \rho_+}}\r].   
\er

On combining the contributions due to the positive and negative 
angular momentum states' to the non-super-radiant modes, we obtain 
the corresponding entropy (evaluated at the Hawking temperature) 
to be
\beq
{\cal S}_{\rm NS} 
= \l(\frac{\lambda \zeta(4)}{8 \pi^3}\r)\, 
\int\limits_{L_{1}}^{L_{2}}\,  \frac{dr}{r\, P}\, 
\l[-\rho_{+} - \frac{3\rho_+^3}{\rho_-^2} 
+ \frac{\rho_+ (\rho_+^2 -\rho_-^2)}
{2 \rho_+\epsilon} + \frac{3 \rho_+^2}{\rho_-} \frac{\sqrt{\rho_+^2 
- \rho_-^2}}{\sqrt{2 \epsilon \rho_+}}\right].
\eeq
Similarly, the entropy associated with the superradiant modes can be
obtained to be
\beq
{\cal S}_{\rm SR} = \l(\frac{\lambda \zeta(4)}{8 \pi^3}\r)\,
\int\limits_{L_{1}}^{L_{2}}\, \frac{dr}{r\, P}\, 
\l[\rho_{+} + \frac{3\rho_{+}^{3}}{\rho_{-}^{2}} 
+ \frac{\rho_{+} (\rho_{+}^{2} -\rho_{-}^{2})}{2 \rho_{+}\epsilon} 
-\frac{3 \rho_{+}^{2}}{\rho_{-}} 
\frac{\sqrt{\rho_{+}^{2} -\rho_{-}^{2}}}{\sqrt{2 \epsilon 
\rho_{+}}}\r]
\eeq
so that the the total entropy of the black string, evaluated at the
Hawking temperature, is given by
\beq
{\cal S}_{\rm BS} 
= \l(\frac{\lambda\, \zeta(4)}{8 \pi^3}\r)\, 
\int\limits_{L_{1}}^{L_{2}}\, \frac{dr}{r\, P}\, 
\l(\frac{\rho_{+}^{2}-\rho_{-}^{2}}{\epsilon}\r)
\eeq
On subsituting the expression~(\ref{eq:Nrbs}) for $N$ in the 
definition~(\ref{eq:ico}),  we find that, up to the leading 
order in $\epsilon$, the invariant cut-off is given by
\beq
\tilde{\epsilon} 
=\l(\frac{r\, \sqrt{2 \rho_{+} \epsilon}}{\sqrt{\lambda\,
\l(\rho_{+}^{2}-\rho_{-}^{2}\r)}}\r).
\eeq
Therefore, the entropy of the rotating black string can be 
expressed in terms of the invariant cut off as follows:
\beq
{\cal S}_{\rm BS}
= \l(\frac{\zeta(4)}{4 \pi^3\, {\tilde \epsilon}^{2}}\r)\, 
\l(\rho_{+}\, {\cal L}\r)
=\l(\frac{\zeta(4)}{8\, \pi^4\, 
{\tilde \epsilon}^{2}}\r)\, {\cal A}_{\rm BS}
\eeq
where ${\cal L}$ is given by Eq. (\ref{eq:calL}) and 
${\cal A}_{\rm BS}=\l[(2\pi\, \rho_{+})\, {\cal L}\r]$ denotes 
the area of the rotating string.
Note that the integral over the extra dimensional bulk is same 
as in the non-rotating case.
Moreover, in the non-rotating limit, $\rho_{+}$ reduces to 
$\rho_{\rm H}$, so that we recover the results we had obtained 
in the previous section. 

\section{Discussion}\label{sec:dscssn}

In this work, we have evaluated the entropy of BTZ black strings 
that are confined to a two-brane in a four dimensional anti-de 
Sitter bulk using the brick-wall approach.
We find that the Bekenstein-Hawking `area' law is satisfied both 
on the brane as well as in the bulk.
It will be worthwhile to construct a generic proof that shows that 
the contribution due to the bulk modes does not affect the 
entropy-area relation for an an arbitrary black string (in this
context, see, for example, Refs.~\cite{kenmoku06}).    
We hope to discuss this issue in a future publication.

\begin{acknowledgments}
The authors would like to thank S.~Shankaranarayanan for comments on 
an earlier version of the manuscript and A.~J.~M.~Medved for helpful 
correspondence.
HKJ wishes to thank T.~Padmanabhan for useful discussions. 
\end{acknowledgments}

\end{document}